# Going Ballistic: Graphene Hot Electron Transistors

S. Vaziri[1], A.D. Smith[1], M. Östling[1], G. Lupina[2], J. Dabrowski[2], G. Lippert[2], F. Driussi[3], S. Venica[3], V. Di Lecce[4], A. Gnudi[4], M. König[5], G. Ruhl[5], M. Belete[6], M.C. Lemme[6,*]

[1]*KTH Royal Institute of Technology, School of Information and Communication Technology, Isafjordsgatan 22, 16440 Kista, Sweden*

[2]*IHP, Im Technologiepark 25, 15236 Frankfurt (Oder), Germany*

[3]*University of Udine, via delle Scienze 208, 33100 Udine, Italy*

[4] *E. De Castro Advanced Research Center on Electronic Systems, University of Bologna, Bologna 40135, Italy*

[5]*Infineon Technologies AG, Wernerwerkstrasse 2, 93049 Regensburg, Germany*

[6]*University of Siegen, Hölderlinstrasse 3, 57076 Siegen, Germany*

*corresponding author: max.lemme@uni-siegen.de

**Abstract**

This paper reviews the experimental and theoretical state of the art in ballistic hot electron transistors that utilize two-dimensional base contacts made from graphene, i.e. graphene base transistors (GBTs). Early performance predictions that indicated potential for THz operation still hold true today, even with improved models that take non-idealities into account. Experimental results clearly demonstrate the basic functionality, with on/off current switching over several orders of magnitude, but further developments are required to exploit the full potential of the GBT device family. In particular, interfaces between graphene and semiconductors or dielectrics are far from perfect and thus limit experimental device integrity, reliability and performance.

**Introduction**

The experimental realization of graphene [1] and other two-dimensional (2D)-materials [2] has opened up new opportunities for pushing the limits of the state-of-the-art in electronics [3], [4] and photonics [5], [6]. This has been motivated by graphene's excellent material properties, which surpass those of conventional materials in many aspects. Nevertheless, in spite of its high charge carrier mobility [7] and saturation velocity [8], graphene field effect transistors (GFETs) struggle to match or surpass the performance of conventional silicon FETs. Fundamental challenges originate in the electronic band structure of graphene. The absence of a band gap leads to high off-state currents and low on/off current ratios, which prohibit GFET applications as logic gates [9][10]. Another consequence of the zero band gap is band to band tunneling, which reduces the output current saturation and the voltage gain, limiting the RF performance potential of GFETs [11], [12]. Recently, vertical electronic device concepts have been proposed to overcome this intrinsic limitation [13]–[19]. One of these novel device concepts, introduced by Mehr et al. in 2012, is vertical graphene base transistor (GBT) [13]. The concept of the GBT is based on the metal-base hot-electron transistors (HETs) introduced originally by Mead in 1961 [20]. HETs utilize high energy tunneling injected electrons (hot electrons) to reach high performance [21]. The first HETs were composed of metal emitters, metal bases, and metal collectors, which were isolated from each other by thin oxide layers. One of the main challenges for the HETs as well as heterojunction bipolar transistors (HBTs) is that the cutoff frequency is limited by base transit time. While thinning down the base mitigates this issue, it dramatically increases the base resistance, resulting in high RC delay and self-bias crowding. The graphene base transistor, in contrast, exploits the high conductivity and the single atomic-thinness of graphene as the base material in HETs to minimize the base transit time and achieve high cutoff frequencies. This concept is distinctly different from vertical graphene field effect tunneling transistors as introduced by Britnell et al. [16]. While the latter functions due to the limited density of states in single layer graphene and electrostatic gate control of the carrier transport between two isolated single layer graphene (SLG) sheets, the GBT operates through emitter-base barrier modulation analogous to the bipolar technology,. This article reviews the experimental and theoretical progress on the GBTs and related devices.

**Working principles of the GBT**

The difference between the GBT and the GFET is shown schematically in Figure 1. In the GFET, carrier transport happens in the graphene plane between the source and the drain with a $V_{ds}$ bias, while the gate electrostatically controls the conductivity of the graphene channel (Figure 1a). In the GBT, in contrast, carriers move perpendicular through the graphene plane. The graphene base is isolated from a metal or doped semiconductor emitter and collector by an emitter-base insulator (EBI) and a base-collector insulator (BCI). Figure 1c and 1d illustrate the simplified band diagram of a GBT in the off-state and the on-state, respectively. The GBT collector current can be modulated by the emitter-base voltage, if an appropriate, fixed emitter-collector voltage is applied (common emitter configuration) and if an appropriate electron barrier height and thickness is chosen for the EBI. When the emitter-base voltage is low, electrons cannot be injected and the device is in the off-state (Fig. 1c). When the emitter-base voltage is high, the effective barrier thickness is reduced, enabling electron injection to the graphene base through Fowler-Nordheim tunneling and onwards towards the collector (Figure 1d). Injected electrons with energies comparable to the emitter Fermi level are considered as hot electrons. Finally, by choosing a low BCI barrier to suppress or minimize quantum mechanical backscattering phenomena at the base-collector interface, the injected hot electrons contribute to the collector on-current, ideally approaching a current gain of 1. Alternatively, EBIs with low barrier heights can facilitate the emission of electrons by thermionic emission (not shown) as a result of effective barrier height lowering. Due to the 2D nature of graphene, hot electron motion through the atomically thin material could approach ideal ballistic transport – resulting in quasi-zero base transit time.

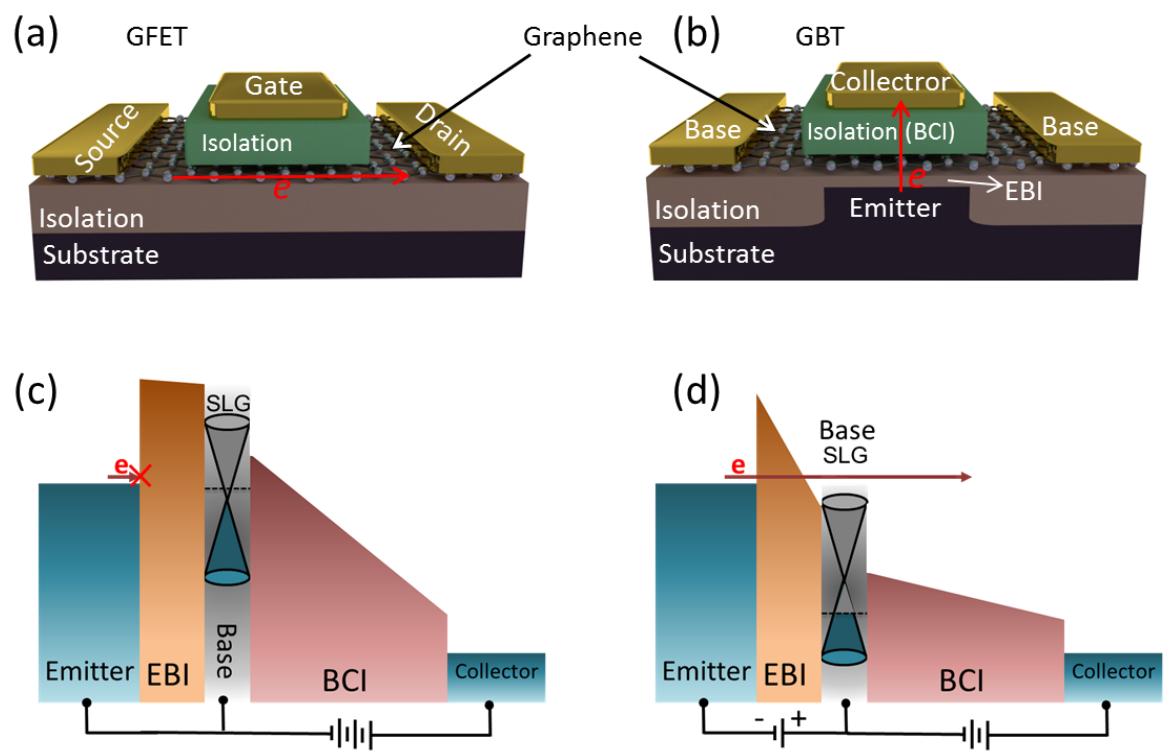

**Figure 1 Schematics of (a) GFET and (b) GBT. The red arrow shows the direction of electron transport in the on-state of these devices. (c,d) Simplified band diagram of the GBT in the (c) off-state and (d) on-state in the common-emitter configuration. We note that in (c) and (d) the**

**energy difference between Fermi level and Dirac potential in graphene represents graphene's quantum capacitance effect.**

The performance of GBTs strongly depends on the design parameters to maximize the current and minimize the loss mechanisms. Thanks to the one-atom thick graphene, the scattering of hot electrons in the base is already minimized. However, EBI parameters need to be accurately chosen to guarantee high injection current densities. Simultaneously, the EBI needs to prevent the emission of cold electrons with energies comparable to the base Fermi energy via defect mediated electron transport or direct tunneling. These cold electrons are not able to surpass the base-collector barrier and lead to the parasitic base current. Cold electron transport limits the common-base current gain or current transfer ratio α ($\frac{I_C}{I_E}$). Furthermore, the BCI needs to act as an electron filter, which allows the passage of the hot electrons and blocks the cold electron emission from the base to the collector. This requires a low barrier to minimize the quantum mechanical backscattering of hot electrons at the BCI barrier, and, simultaneously, suppress thermionic, tunneling, and defect mediated electron transport from base to collector. Consequently, modeling of GBTs to define a window of optimized design parameters for high performance operation is essential.

**Device modeling/simulation and performance projection**

A zero-order estimation of the GBT performance has been performed based on quantum-mechanical simulations in [13]. For that purpose, the Schrödinger equation with open-boundary conditions was solved numerically for one-band effective potential rounded up by image force at interfaces with emitter and collector. This early model, with no scattering effects included, predicted that for a terahertz operation at emitter-base voltages around 1V, an EBI with a barrier of 0.4 eV or smaller and a thickness lower than 3-5 nm is required. Fig. 2 shows simulated transfer and output characteristics for a device with $Er_2Ge_3$/Ge emitter ($\Phi_{EBI}$ = 0.2 eV) and a compositionally graded $Ti_xSi_{1-x}O_2$ BCI. Clearly, the device shows switching over several orders of magnitude and saturating output characteristics.

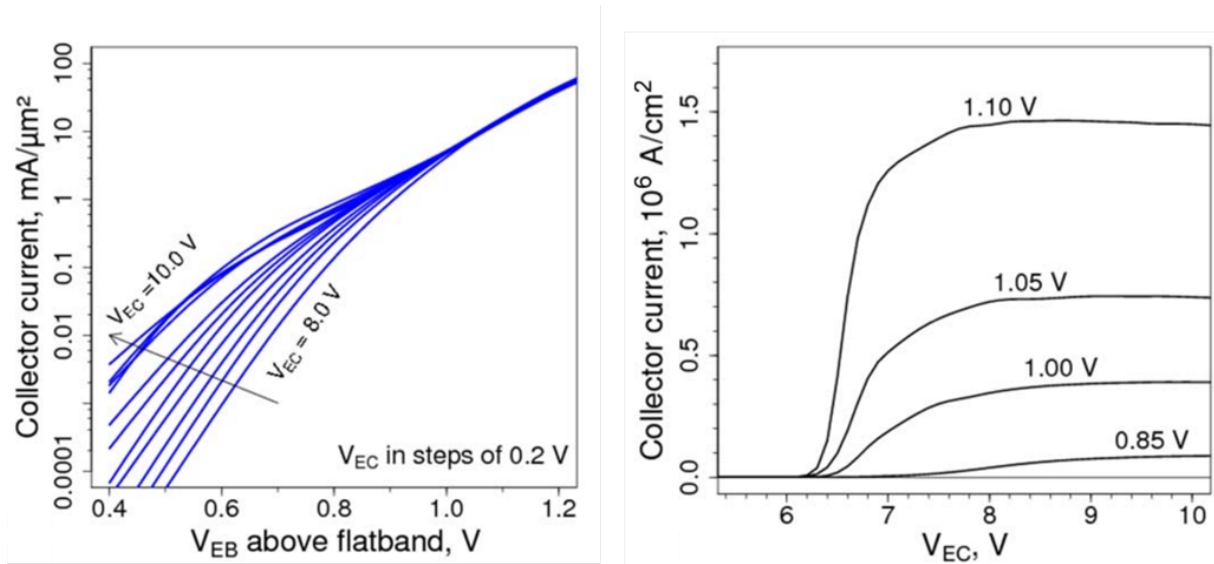

**Figure 2** Initial simulation results. Transfer (a) and output (b) characteristics of a GBT with 3nm EBI and 80nm BCI.

More recent modeling implementations have confirmed the potential of the GBT and explored the design space of the device [22]. Venica et al. have proposed a model that calculates the GBT electrostatics self-consistently with the charge stored in the graphene and the electrons tunneling through the EBI and BCI [23]. In this way, the model accounts for the electrostatic impact of the charge traveling along the GBT. As a result, space charge effects at high current levels (that usually reduce the maximum $f_T$ in bipolar transistors) are considered. Since the physical origin of the base current is still unclear and debated [15], the model assumes a priori a negligible base current and the collector current density ($J_C$) is thus due to the electrons injected from the emitter.

The model calculates the I-V characteristics and it has been verified through comparison with available experiments ($SiO_2$ EBI in Fig. 9a). In addition, it estimates $f_T$ by means of a quasi-static approach and the unity power gain frequency ($f_{max}$), by considering a base resistance ($R_B$) as the sum of the intrinsic graphene resistance ($R_{INT}$) and the contact resistance between the graphene layer and the base contact ($R_{CONT}$) [24]. The model confirmed the possibility of the GBT's potential for THz operation within a fairly large design space. In particular, Fig. 3a reports on the $f_T$ vs. $J_C$ curves of an optimized GBT with Si emitter, $Ta_2O_5$ EBI and SiCOH BCI [25]. By comparing symbols and dashed lines in Fig. 3a, we note that space charge effects are present also in GBTs, but are less important than in bipolar transistors. Furthermore, by an appropriate optimization of the GBT geometry, it is possible to limit the detrimental effect of the contact resistance between the metal and the graphene and to obtain large $f_{max}$ values (see Fig. 3b).

Fig. 4a shows the output characteristics of the GBT shown in Fig. 3. The curves show quite good saturation, indicating that GBTs can overcome the GFET limitations concerning the output conductance ($g_d$). As a result, the predicted intrinsic gain $g_m/g_d$ reaches values up to 50 (Fig. 4b).

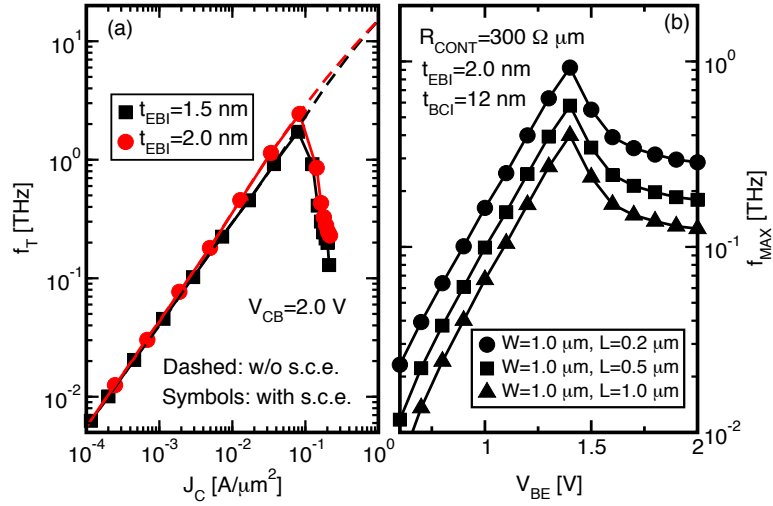

Figure 3 (a) Cutoff frequency versus collector current for an optimized GBT exploiting Si emitter, $Ta_2O_5$ EBI and SiCOH BCI [25]. Although space charge effects (s.c.e.) in BCI limit $f_T$ at high current (compare symbols with dashed lines), THz operation is still feasible. (b) $f_{max}$ versus $V_{BE}$ for different GBT geometries. We assumed $R_{CONT}$=300 Ω.µm. The device is contacted at both sides. Optimization of dimensions allows $f_{max}$ around THz.

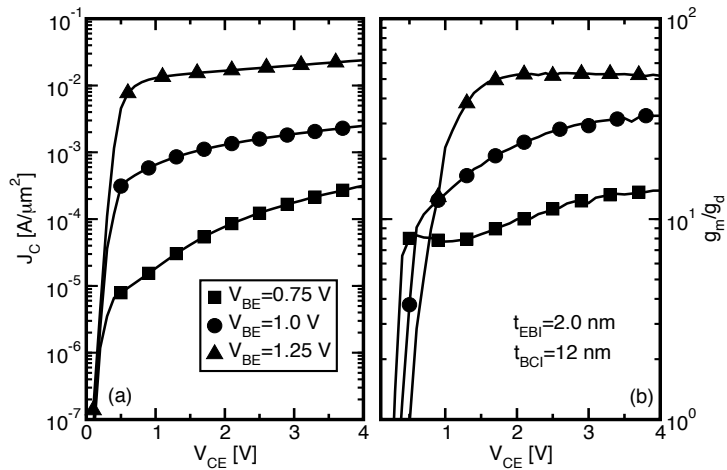

Figure 4 (a) JC vs. $V_{CE}$ of the GBT of Fig. 3 showing the quite good saturation of the output characteristics. (b) The good intrinsic gain of the GBT.

Di Lecce et al. have performed a simulation study of GBTs with semiconducting EBI and BCI layers with an self developed tool based on a full-quantum mechanical transport approach coupled with Poisson equation [26]. A 1-D device model was assumed, with uniformity in the transverse directions, ignoring transport through the valence bands of the semiconducting layers. Electron transport within the semiconducting layers was solved within the ballistic non-equilibrium Green's function (NEGF) formalism using an effective mass Hamiltonian. In order to correctly estimate the group velocity of high energy electrons, in particular of those in the BCI region near the collector, non-parabolic corrections to the effective mass model were introduced, adopting the Flietner's energy dispersion relation [27] for the conduction-band valleys with a non-parabolicity coefficient α=0.5 eV$^{-1}$. Space charge effects due to traveling electrons, band structure effects in the semiconducting layers such as multiple ellipsoidal valleys, and the effect of graphene on the device electrostatics were fully accounted for. Two semi-infinite leads mimicked the emitter/collector contacts.

Two reference devices, GBT1 and GBT2, were simulated, with a BCI layer thickness $t_{BCI}$ of 20 nm and 10 nm, respectively. The emitter and collector Schottky barriers were set to 0.2 eV and the EBI layer thickness $t_{EBI}$ to 3 nm.

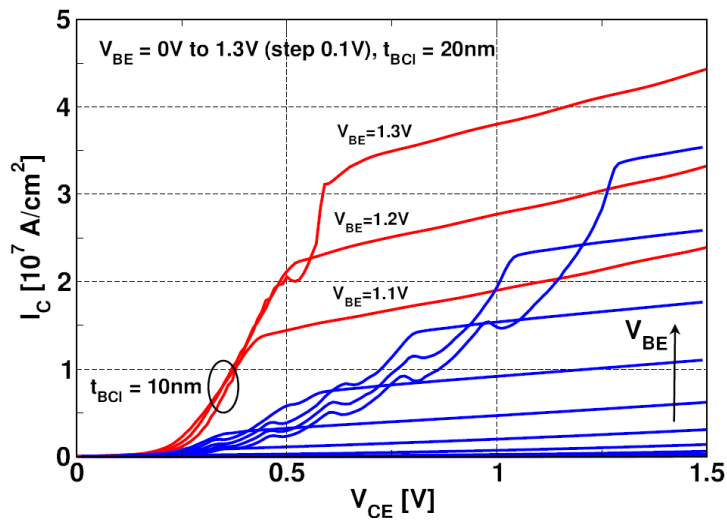

**Figure 5 Output characteristics of the device GBT1 for $V_{BE}$ ranging from 0 to 1.3 V and GBT2 for $V_{BE}$ ranging from 1.1 to 1.3 V, both in steps of 0.1 V. The negative differential resistance at low $V_{CE}$ is believed to be related to quantum resonance effects in the potential well surrounding graphene.**

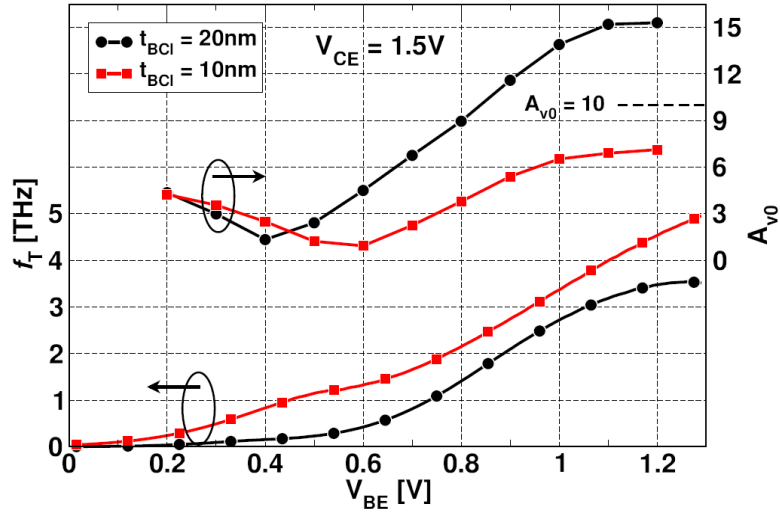

Figure 6 Cut-off frequency (left axis) and intrinsic voltage gain (right axis) for devices GBT1 and GBT2 versus $V_{BE}$ for $V_{CE}$ = 1.5 V

The output characteristics show that the current is in the range of $10^7$ A/cm$^2$ even if it is due to tunneling through the EBI layer (Fig. 5). Device GBT1 shows a better output conductance $g_{CE}$ in the saturation region, but a higher saturation voltage, which represents a potential drawback. In [26], the unsaturated current region (low $V_{CE}$) was associated with the presence of a potential barrier within the BCI layer, sustained by the negative space charge due to the traveling electrons, which retards the onset of saturation with increasing $V_{CE}$. The larger $g_{CE}$ of GBT2 can be explained with the higher electrostatic effect of the collector voltage on the graphene charge and Dirac voltage, hence on the EBI tunneling length, due to the thinner BCI layer compared with GBT1.

The intrinsic voltage gain $A_{v0}$ and the unit short-circuit current-gain frequency $f_T$ are reported in Figure 6. As can be seen, well-above-THz operation is within reach for both GBT1 and GBT2; the lower $t_{BCI}$ results in a higher $f_T$, but, due to the higher output conductance, the $A_{v0}$ is worse. The dip in the $A_{v0}$ curve around $V_{BE}$ = 0.4-0.6 V is due to the limited quantum capacitance $C_Q$ of graphene, which adversely affects the device transconductance. On the contrary, $f_T$ exhibits a monotonically increasing behavior and is not affected by $C_Q$.

**Proof-of-concept and experimental realization of the GBT**

In 2013, Vaziri et al. demonstrated the first experimental proof of concept GBT on a chip / die scale [14]. The device comprised of an n-doped silicon emitter, a 5 nm-thick thermal silicon dioxide ($SiO_2$) EBI tunneling barrier, a graphene base, a 15-25 nm-thick atomic layer deposited (ALD) aluminum oxide ($Al_2O_3$) BCI, and a metal (titanium/gold) collector. The fabrication was done on 8-inch wafers and the GBTs were isolated by 400 nm $SiO_2$ shallow trench isolation (STI) making the fabrication scheme potentially CMOS compatible [28]. The reported DC functionality of this device confirmed the working principles of the GBT. Figure 5a illustrates the transfer characteristics of a GBT at $V_{BC}$ of 2 V. The device is in its off-state before the onset of the collector current at an emitter-base voltage of approximately 4.5 V, i.e. the threshold voltage. Moreover, thanks to the good isolation characteristics of 20 nm $Al_2O_3$ BCI, the device shows very low off-state currents and more than four orders of magnitude on/off ratio. Fig. 7b shows temperature dependent I-V characteristics of the emitter-base tunneling diode (input characteristics). Clearly, the dependence of the current to the temperature diminishes for high voltage range. This fact together with the excellent linear fit of the I-V characteristics to the Fowler-Nordheim model (inset of Fig. 7b) confirm that the transport mechanism is dominated by tunneling. In addition, in Fig. 7b the onset of the emitter-base tunneling current is well matched to the onset voltage of the collector current in Fig. 7a.

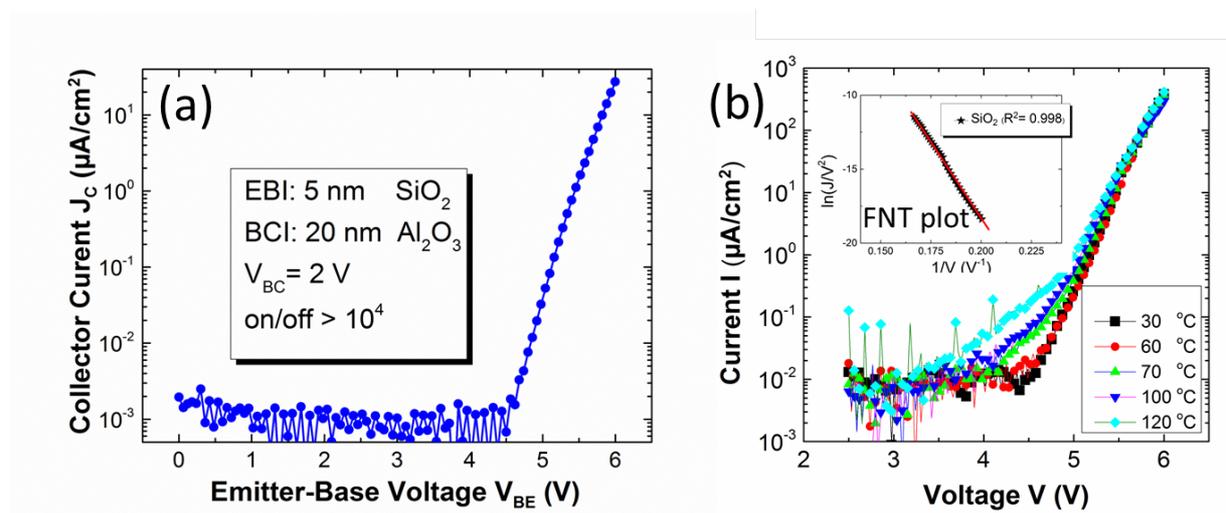

**Figure 7 (a) Transfer characteristics of a GBT with 5 nm thick $SiO_2$ and 20 nm thick $Al_2O_3$ as the EBI and the BCI, respectively. The transfer characteristics shows an on/off ratio exceeding $10^4$. (b) Temperature dependence I-V for the injection tunnel diode with 5 nm $SiO_2$ as the tunnel barrier. The inset FN plot shows an excellent linear fit.**

Fig. 8a shows the output characteristics of a GBT in the common-emitter configuration ($V_E=0$), in which $I_C$ is measured as a function of $V_C$ at different base voltages. For $V_{BE} = 3$ V, there is no injection from the emitter (Fig.7b) resulting in zero collector current (i.e. at the noise floor of the measurement setup). However, when electron injection starts for base voltages greater than 4.5 V, these electrons can pass through the device to reach the collector only when the collector barrier is lowered enough ($V_{EC} > 3$ V). This is illustrated more clearly in the logarithmic-scale output characteristics shown in the inset of Fig. 8a. At higher $V_{BE}$, larger collector current densities are obtained. The common-base output characteristics of another GBT with the identical device parameters as in Fig. 8a is shown in Fig. 8b. In this configuration, the base is biased at 0 V, current is forced to the emitter, and the collector current is measured as the collector voltage is swept from 0 to 4 V. Moreover, the inset shows the emitter voltage feedback as a function of the collector voltage. When the emitter current is zero, no collector current is observed. For higher emitter currents, in contrast, the device is in the on-state (red circles and blue triangles).

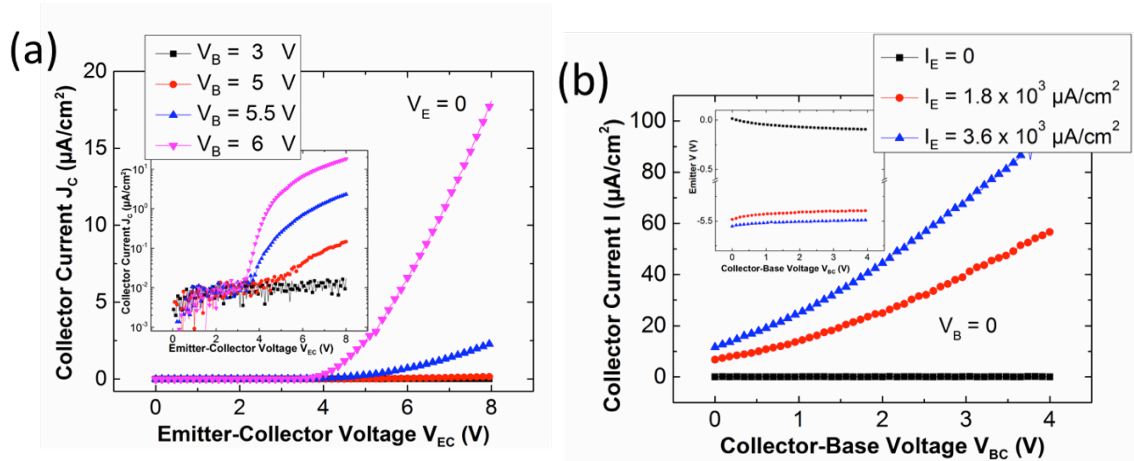

**Figure 8 (a) Common-emitter output characteristics for various base voltages $V_B$. The inset shows the logarithmic scale of the same graph. (b) common-base output characteristics at three different forced emitter current densities. The emitter voltage feedback is illustrated in the inset.**

While the proof of concept devices confirmed the DC functionality of the GBT, the performance did not reach the predictions due to non-optimized design parameters. Specifically, the devices showed low collector current density and a limited current transfer ratio α (or common-base current gain) of 6.5%, preventing efficient high frequency performance. The limitations originate from the characteristics of the 5 nm $SiO_2$ tunneling barrier and the $Al_2O_3$ BCI. While these materials were chosen for their well-known behavior and high quality for the proof of concept devices, their characteristics are far from the criteria needed for a high performance GBT. Considering the exponential dependence of the tunneling current on the barrier height and thickness, a 5 nm $SiO_2$ EBI with a 3.3 eV barrier height with respect to the conduction band of silicon [29], [30] dramatically limits the injected emitter current. On the other hand, $Al_2O_3$ forms a 3.4 eV barrier height with respect to the graphene Fermi level [31] reducing the collector current and the current transfer ratio due to the large quantum mechanical backscattering. As a consequence, it is expected that optimized tunneling and filtering barrier heights and thicknesses will dramatically improve the device performance. Moreover, efficient transfer of high quality pinhole-free graphene layers, low metal-graphene contact resistance, and high quality EBI and BCI materials with good interfaces are critical to achieve high performance GBTs.

Zeng et al. reported functional GBTs with clear saturation behavior of the output characteristics and improved current transfer ratio α [15]. Their work includes GBTs with four different sets of material parameters showing how the transfer ratio can be affected by these design parameters. For example, in two devices with the same EBI of 25 nm $SiO_2$, a four times improvement in the current transfer ratio was reported using 21 nm of hafnium dioxide ($HfO_2$) as the BCI instead of $Al_2O_3$ with the same thickness. This can be attributed to the 1.3 eV lower band offset of the $HfO_2$. While the maximum current transfer ratio α was about 5%, the effective α (normalized to the effective collector area) was about 40%. However, besides very low currents, the devices with higher α values exhibited low on/off ratios. This again emphasizes the importance of the device layout and material parameters in the GBT device performance.

**Optimizing the Barriers**

**Emitter-Base-Insulator (EBI)**

High frequency performance of the GBT requires high on-state collector currents $I_C$. In an ideal device with α=1, $I_C$ is equal to the emitter current $I_E$, which consists of injected hot electrons. Hence, the EBI barrier should be low and thin enough to provide a high current of hot electrons, yet block cold electron emission. To this end, a number of potential materials have been investigated. For example, replacing a 5 nm $SiO_2$ EBI with 6 nm thick ALD $HfO_2$ results in an improved threshold voltage and a higher emitter current (Fig. 9a). This 6 nm thick insulator includes a 0.5 nm interfacial $SiO_2$ to improve the Si/dielectric interface quality. However, the choice of dielectric materials to form a low band offset barrier is limited to the high-k dielectrics, which are known for reduced film and interface integrity compared to $SiO_2$. This means that for thin tunneling layers, the defect mediated carrier transport or unfavorable direct tunneling could become the dominant transport mechanisms. One possible solution are bilayer dielectric tunnel barriers, where a first thin layer in contact with the emitter provides a good interface and high material quality, while a second layer reduces off-state leakage. Together, the bilayer dielectric minimizes defect mediated carrier transport. This approach facilitates the application of low band gap (high electron affinity) dielectrics to improve the emitter current by promoting FNT or step tunneling [32] (Figure 9b). Figure 9a shows dramatic improvement in the emitter current by applying a novel $TmSiO/TiO_2$ (1 nm/5 nm) bilayer tunnel barrier. We found that high electron injection through this dielectric stack is dominated by tunneling [33].

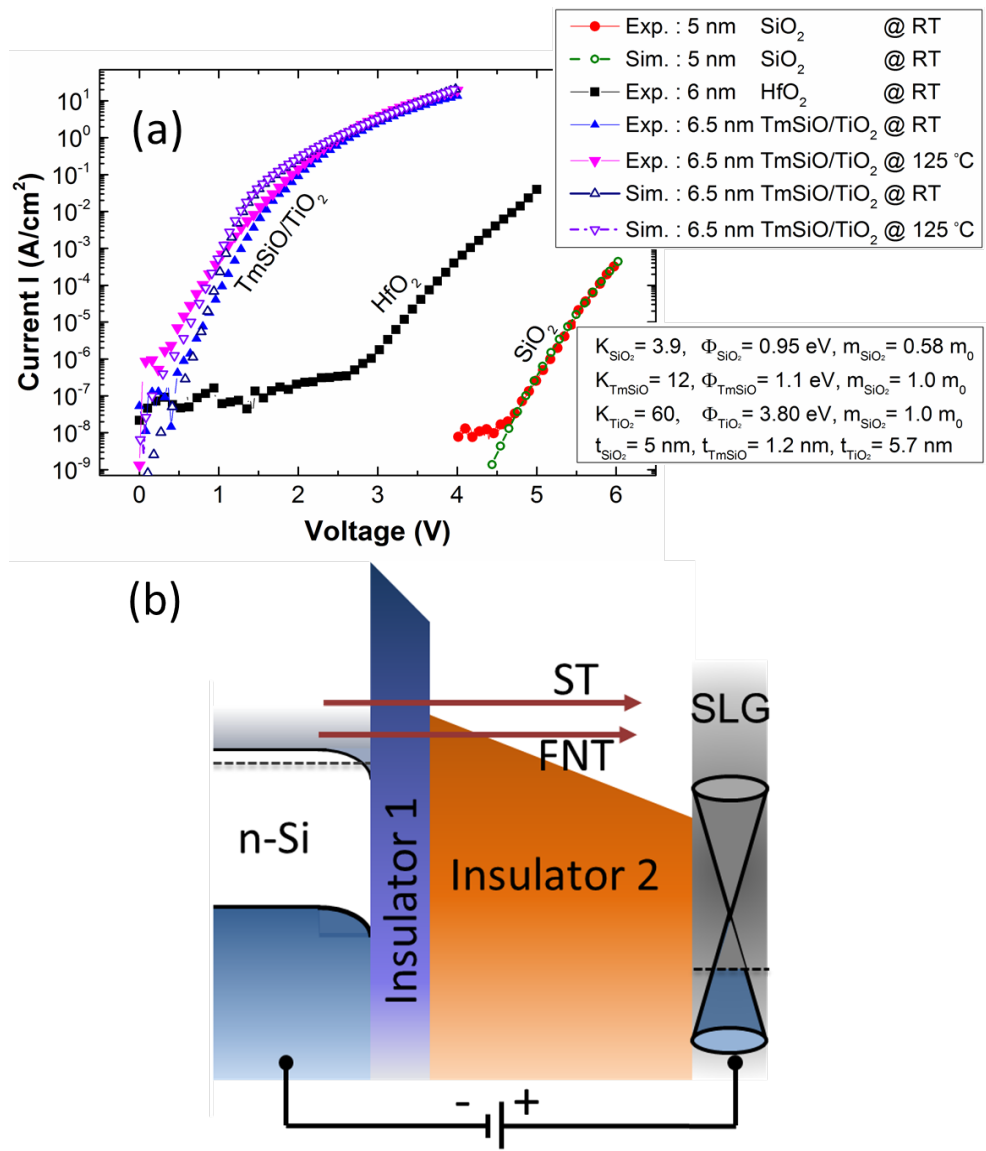

**Figure 9 (a) I-V characteristics of Si-insulator-graphene tunnel diodes with $SiO_2$, $HfO_2$, and $TmSiO/TiO_2$ tunnel barriers. The $HfO_2$ tunnel barrier with lower band offset with respect to Si conduction band shows improved I-V characteristics (blue triangulars) in comparison to $SiO_2$ (brown squares). Superior I-V characteristics have been achieved using $TmSiO/TiO_2$ bilayer tunnel barrier (red circles). The tunneling transport mechanisms in the bilayer tunnel barriers are shown in the (b) simplified band diagram.**

Other structures to improve the emitter-injected current have also been proposed. The carrier transport in graphene on GaN/AlGaN heterostructure has been studied for both potential applications in graphene vertical tunneling field effect transistors and GBTs [34], [35]. This heterostructure provides a 2DEG at the GaN/AlGaN junction as the emitter and a low barrier with crystalline quality, which ensures high thermionic electron emission. Alternative structures utilizing 2D crystalline materials and bulk semiconductor heterojunctions have been also proposed [18], [19]. These structures will be briefly discussed in the following sections.

**Base-Collector-Insulator (BCI)**

The devices discussed thus far were based on the formation of the EBI on the wafer or substrate prior to graphene transfer onto the EBI. This is due to graphene's inert surface, which makes it very challenging to deposit high quality thin dielectric/semiconductor layers on top of graphene. On the other hand, the BCI can be much thicker than the EBI, and can therefore be formed reliably on top of the graphene. In previous reports, the formation of the BCI on graphene has been accomplished by evaporation of a 2-3 nm-thick aluminum seed layer and consecutive atomic layer deposition of $Al_2O_3$ or $HfO_2$. There are several concerns in this technology. First, uncompleted oxidation of the seed layer can introduce charge trap sites and fixed charges to the graphene-dielectric interface. Moreover, this layer may dramatically affect the effective barrier height of the BCI. Another issue is that materials thus far utilized as BCIs, i.e. $Al_2O_3$ and $HfO_2$, form rather large barrier heights with respect to the graphene Fermi level. Finally, experimentally realizing low defect and thin isolation layers of low band gap dielectrics like $TiO_2$ and $Ta_2O_5$ is not straightforward.

One promising approach could be the deposition of semiconductor materials on graphene as the BCI such that the graphene-semiconductor junction forms a Schottky barrier. A low Schottky barrier could ideally minimize the reflection of hot electrons at the base-collector interface. The formation of thin and conformal Si films on graphene using standard recipes established in Si device manufacturing is, however, also very challenging. This is mainly due to the hydrophobic surface of graphene providing no nucleation sites for chemical vapor deposition (CVD) or ALD processes. Accordingly, silicon (Si) deposition experiments using standard CVD epitaxy recipes did not yield satisfying results. For example, experiments with disilane precursor on transferred graphene resulted in the growth of Si islands which did not form a closed layer even at a nominal thickness of about 100 nm. Significantly better results were obtained with physical vapor deposition (PVD) by evaporation of silicon using e-beam (Fig. 10). We demonstrated that this method enables the deposition of closed Si layers with low roughness on CVD graphene [36]. At the same time, the crystalline quality of graphene as measured by Raman spectroscopy was largely unaffected. Furthermore, we found that very high frequency plasma-enhanced chemical vapor deposition (PE-CVD) [37] is a suitable tool for the deposition of dense and conformal amorphous Si films as thin as 30 nm. As a result of the very gentle nature of high frequency plasma, no deterioration of the graphene quality was observed during the growth process.

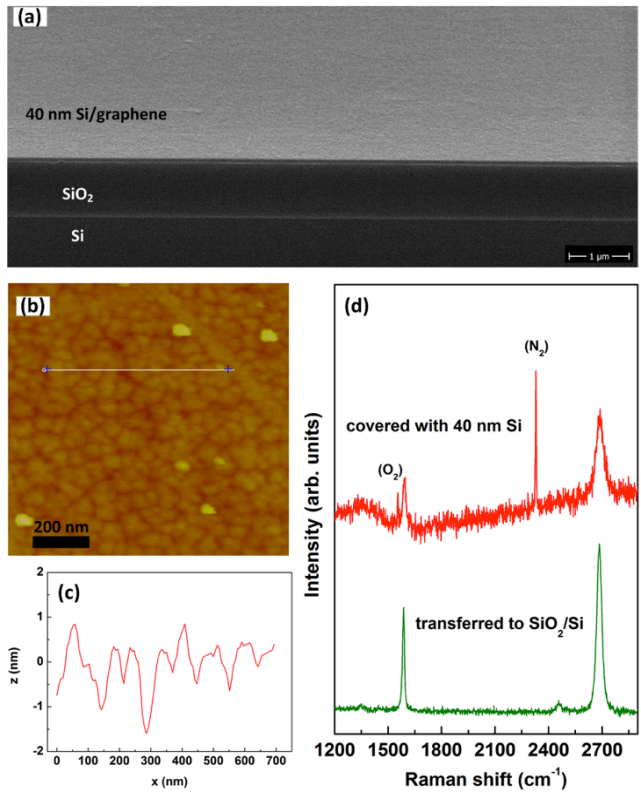

**Figur 10** PVD deposition of Si layers on transferred CVD graphene. (a) tilted view SEM image of a 40 nm Si layer on graphene. (b) Corresponding AFM image. (c) AFM section along the line in (b). RMS roughness is 0.6nm. (d) Raman spectra of graphene acquired before and after the deposition process [36].

**Wafer-scale integration of vertical graphene base transistors**

In parallel to the optimization of GBTs on a chip / die scale, experiments targeting their implementation in a 200 mm wafer Si pilot line were initiated. A wet graphene transfer method was adopted to cover areas of up to 20x20 mm$^2$ on pre-patterned 200 mm wafers. Subsequently, high-k dielectrics (e.g. $HfO_2$) were deposited directly onto the CVD graphene using atomic vapor deposition. These process modules were then combined with a standard Al-based back end of the line metallization to provide a complete process flow for GBT fabrication in a Si pilot line. Fig. 11a shows a 200 mm wafer with fabricated arrays of discrete GBTs. Fig. 11b shows a single device with emitter, base, and collector terminals arranged in ground-signal-ground configuration. Fig. 11c presents a bright-field STEM cross-section image outlining the structure of the device. The magnified part is a TEM-EDX image of the base contact region with visible edge of the graphene layer. Here, contacting graphene with TiN resulted in very high contact resistances (several kOhm*µm). However, low ohmic contacts with CMOS compatible materials is among the biggest general challenges for graphene, and beyond the scope of this review. Nickel metallization combined with contact engineering [38] may be seen as a potential solution.

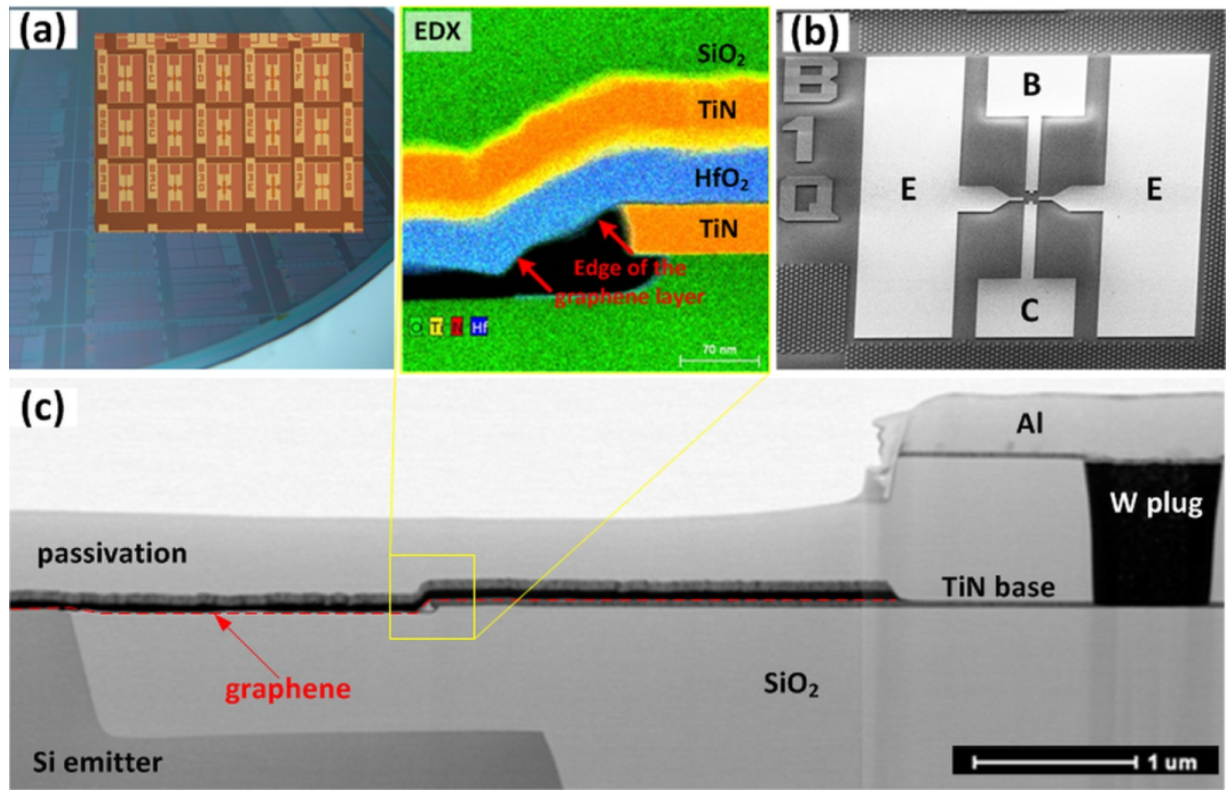

**Figure 11 (a) 200 mm wafer with fabricated GBTs. Inset shows optical microscope image of a matrix of devices with different active areas. (b) SEM image a single GBT with indicated emitter (E), base (B), and collector (C) terminals. (c) Bright-field STEM cross section of the device shown in b. The magnified region shows an EDX image at the edge of the TiN base-graphene contact region. The $HfO_2$ BCI, Ti/TiN collector metal, and the $SiO_2$ passivation are also visible.**

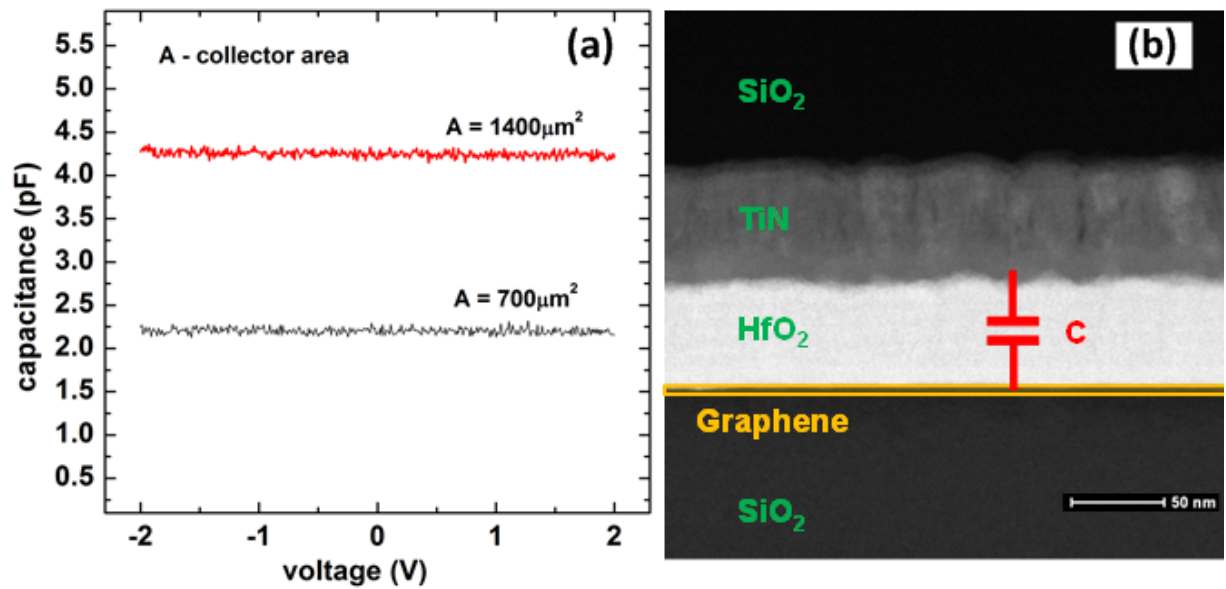

Figur 12 (a) capacitance-voltage curves of BCI from integrated large active area GBT devices on 200 mm wafers. (b) TEM of the graphene-HfO$_2$-TiN capacitor.

The integration of the GBTs on 200 mm wafers revealed that graphene can withstand many steps of harsh technological processing with good quality and low resistivity as shown by Raman and electrical measurements on TLM structures. This was confirmed by electrical characterization of high-k capacitors with graphene electrodes. Fig. 12a shows an example of CV measurements on a HfO$_2$-based BCI embedded between graphene and TiN electrodes [39]. Fig. 12b illustrates an SEM cross sectional image of the measured capacitor. Nevertheless, insufficient purity of transferred CVD graphene was identified as a severe roadblock to further integration of graphene devices in a CMOS pilot line environment. Standard transfer and cleaning protocols result in a relatively large concentration of metallic impurities and thus fail to provide material complying with stringent purity standards of Si fabs [40]. Fig. 13a summarizes TXRF measurements on graphene samples transferred onto 200 mm wafers. Despite careful transfer and extensive cleaning procedures in a large experimental dataset, surface concentrations of metallic impurities below $1 \times 10^{13}$ atoms/cm$^2$ could not be obtained [41]. To investigate the potential influence of such residual metals on the minority carrier diffusion length in the Si substrate, two pieces of graphene from different suppliers were transferred onto p-type Si(100) wafer covered with native SiO$_2$. After transfer, the wafers with graphene were annealed at 600°C for 5 min in N$_2$. Subsequently, the carrier diffusion length was measured point by point to create a map of the wafer (Fig. 13b). These measurements show that the minority carrier diffusion length is significantly reduced in the regions covered with graphene. This reduction is mainly due to transfer related Fe impurities and in a small part due to Cu precipitates [41]. These results indicate a need for further optimization of transfer and cleaning protocols as well as work on alternative metal-free graphene deposition approaches such as growth on Ge [42]–[44].

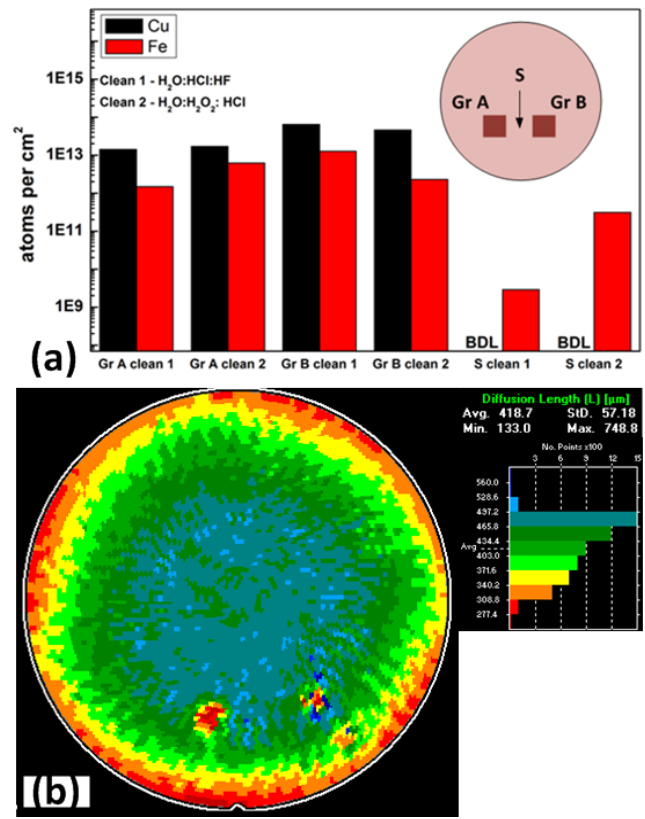

**Figur 13 (a) TXRF measurements on graphene samples transferred to 200 mmm wafers. GrA and GrB stands for two different suppliers of graphene. Clean 1 and Clean 2 stands for two different cleaning protocols. (b) Minority carrier diffusion length measurements on a wafer with transferred graphene samples.**

**Theoretical limits and potential**

The Graphene-Base Heterojunction Transistor (GBHT) proposed by Di Lecce et al. [19] is a promising adaptation of the GBT concept. In this device, graphene is sandwiched between an $n^+$-semiconductor layer (emitter) and an n-semiconductor layer (collector). In this structure the carrier transport mechanism is dominated by thermionic emission over the emitter-base Schottky barrier. As a result, the GBHT is envisioned to overcome part of the GBT's engineering issues, eliminating the need for ultra-thin tunnel barriers and low Schottky barriers. However, as discussed, the formation of a high quality crystalline semiconductor layer on top of graphene remains a processing challenge.

Fig. 14 shows the simplified band diagram of the Si-based GBHT. At zero bias condition, the work function difference between semiconducting emitter/collector and graphene forms depletion regions and band-bending in the both emitter and collector. This band-bending results in a triangular potential energy barrier between the emitter and the collector. In the off-state, in spite of applying a reasonable positive collector-emitter voltage, the triangular barrier blocks the current. However, in the on-state, the height and shape of this barrier can be modulated by the emitter-base voltage. When the height of the barrier is small enough (in the on-state), electrons are injected to the base, pass through graphene and are eventually accelerated by the electric field in the collector region. In terms of structure and behavior, the GBHT is thus similar to an n-p-n HBT, with the p-type base replaced with the graphene monolayer. The model adopted in [19] for the GBHT simulations is the same used for the GBT in [26].

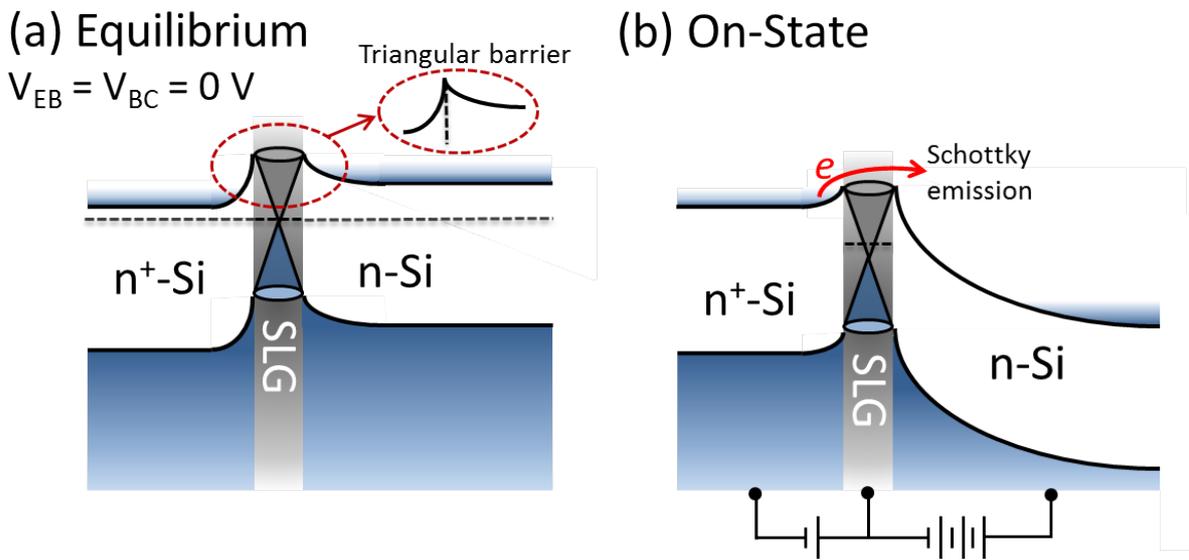

**Figure 14 Band diagram of the GBHT at (a) equilibrium and in the (b) on-state.**

Two reference devices were simulated as case studies, both with an emitter doping concentration of $N_E = 3 \times 10^{19}$ cm$^{-3}$. In the first one, named D1, the collector doping concentration was $N_C = 10^{18}$ cm$^{-3}$, whereas in the second one, named D2, it was $N_C = 3 \times 10^{18}$ cm$^{-3}$. In Fig. 15, the $I_C$–$V_{CE}$ characteristics are reported for different values of $V_{BE}$ for D1 and D2: both are well behaved. An increase in the collector doping (D2) is beneficial to the current values, since the barrier is lower. The $f_T$ and $A_{v0}$ curves are reported in Fig. 16 versus $V_{BE}$: both D1 and D2 show an $f_T$ higher than 1 THz. The better performance of D2 is due to a higher $g_m$ and lower high-injection effects, related to the higher collector doping density. The worse output conductance of D2 (Fig. 15) is responsible for the lower voltage gain, even though it appears to be possible in both devices to simultaneously achieve an $f_T$ higher than 1 THz and a voltage gain higher than 10

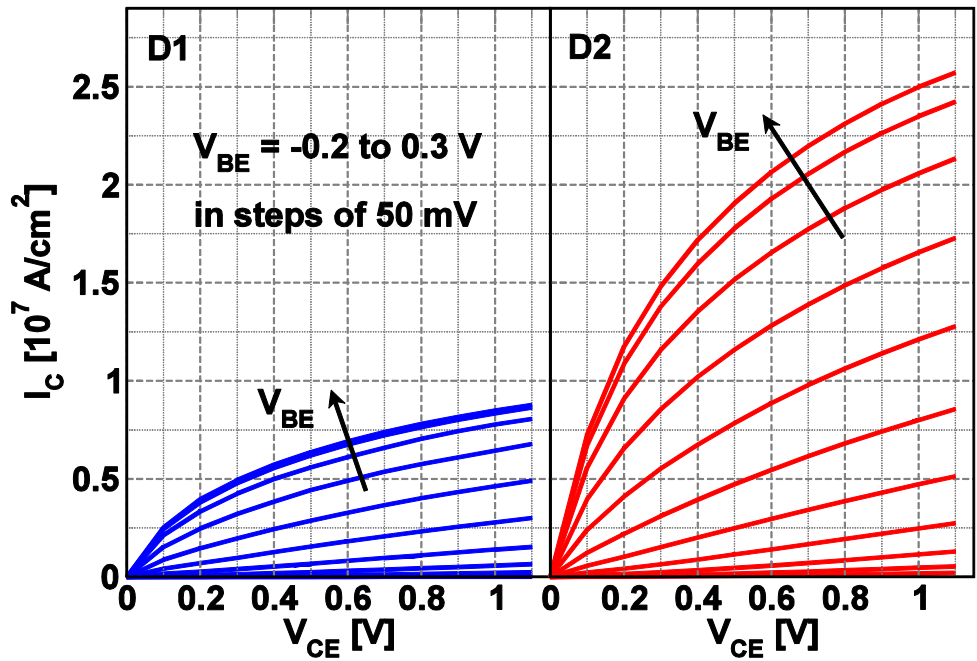

Figure 15 I-V characteristics of GBHTs D1 and D2 defined in the text with $V_{BE}$ sweeping from -0.2 to 0.3 in steps of 0.05 V.

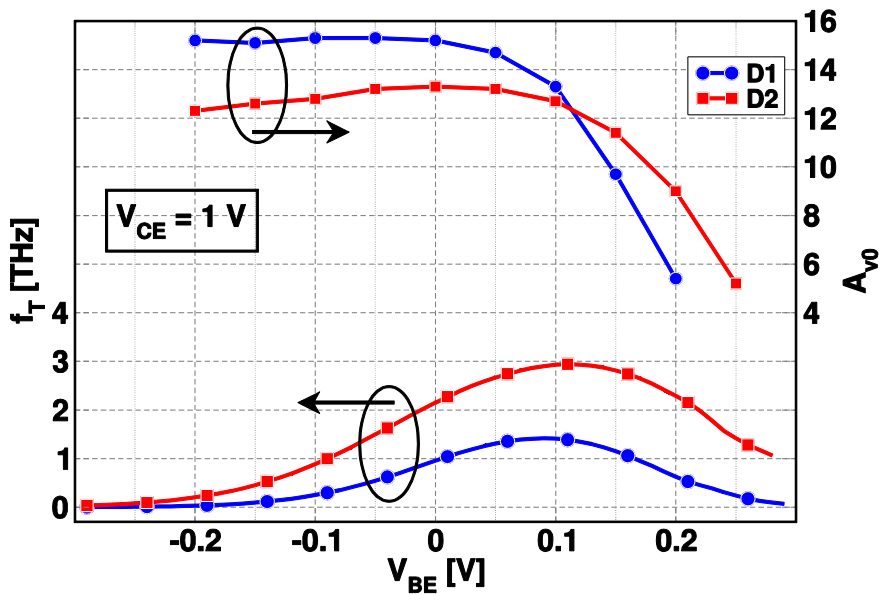

Figure 16 Cut-off frequency (left) and intrinsic voltage gain (right) for GBHTs D1 and D2 defined in the text.

Di Lecce et al. [45] compared also through numerical simulations an n-type Si GBHT and a n-p-n SiGe HBT structure aggressively scaled to reach its performance limits from [46]. The study showed that the GBHT is predicted to deliver an $f_T$ more than twice as high as for optimized HBTs assuming a transparent graphene/Si interface (Fig. 17), thanks to the monolayer thickness of the base region and the smaller emitter transit time due to the unipolar nature of the device (very few holes are present in the n-type device), another key feature which distinguishes the GBHT from the HBT. In Fig. 17c and d are shown the accumulated transit times τ'$_p$(z) and τ'$_n$(z) across the devices at peak f$_T$, defined as $\tau'_p(z) = \int_{z_E}^{z} \frac{d(\rho_{GR}(z') + \rho_p(z'))}{dI_C} dz'$ and $\tau'_n(z) = \int_{z_E}^{z} \frac{d\rho_n(z')}{dI_C} dz'$ with z$_E$ the location of the emitter contact, $\rho_{GR}$ t the charge density on the graphene sheet (set to zero for the HBT), and $\rho_{p/n}$ t the hole/electron charge densities across the devices. The emitter, base and collector transit times for the two devices can be extracted (τ$_E$, τ$_B$, τ$_C$). For the GBHT, τ$_E$ = 6 fs, τ$_B$ = 0, τ$_C$ = 28 fs; for the HBT at V$_{CE}$ = 1 V, τ$_E$ = 25 fs, τB = 55 fs, τ$_C$ = 10 fs.

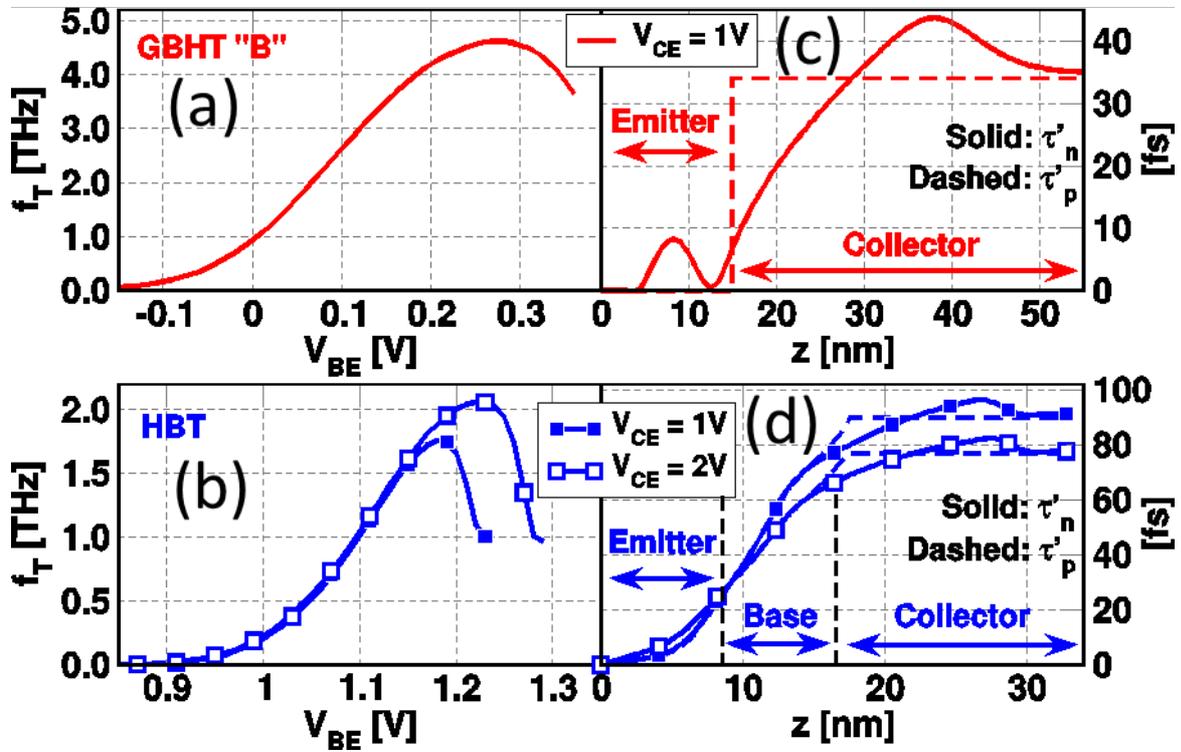

**Figure 17 f$_T$ of the GBHT (a) and HBT (b), showing the capability of the former to outperform the HBT's cut-off frequency by a factor higher than 2. Accumulated transit times τ'$_p$ and τ'$_n$ along the devices, at the peak-f$_T$ bias for the GBHT (c) and HBT (d).**

One potential drawback of the GBHT is the use of highly doped semiconductor layers, with ionized impurities acting as a potentially critical scattering source for electrons. The effect of impurity scattering in the GBHT was investigated in [47], where it was shown that, despite a reduction of the current and cut-off frequency of about a factor 2, the silicon GBHT can still reach $f_T$ values above 1 THz (Figure 18)

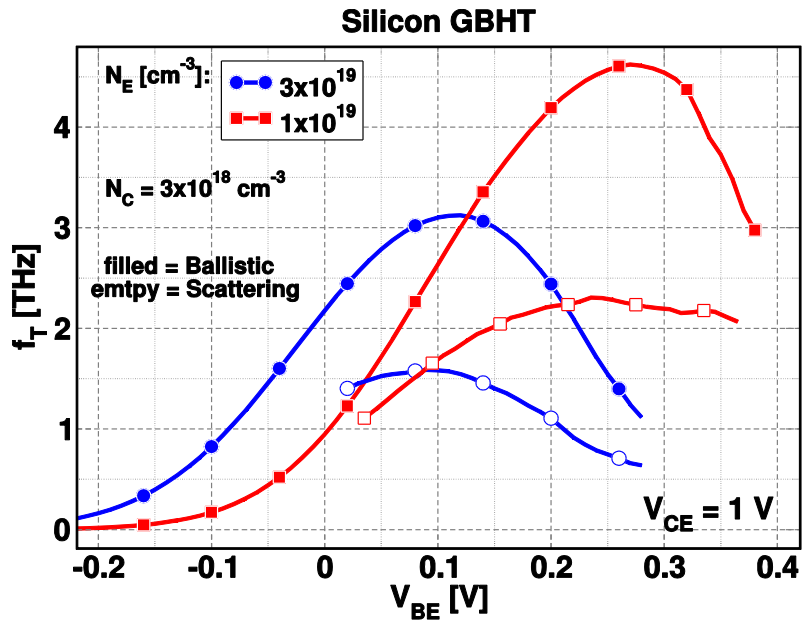

**Figure 18 Simulated cut-off frequency of two different Si GBHT devices in ballistic conditions and with impurity scattering.**

An open critical issue concerning the GBT and the GBHT is the transparency of graphene: an ideal graphene-semiconductor interface was assumed in explorative studies like [26] and [47]. The presence of the graphene base contact can be mimicked by an empirical self-energy $\Sigma_B^R = -\jmath \Delta_B \delta(z - z_B)$ with $z_B$ the location of the graphene base and $\Delta_B$ a coupling constant between the contact and the rest of the device, a fitting parameter to be adjusted on the basis of future experimental base currents, or transport studies of electrons across the graphene layer. In [26] such value was assumed so as to yield an almost ideal $\beta = \frac{I_C}{I_B} \approx 10^5$ in the active region of the GBT. In [45] a non-ideal interface was accounted for in the GBHT by sandwiching graphene between two potential barriers of height $h_B$ and thickness 0.525 nm. With experimental data for GBHTs still unavailable, such barrier height was adjusted together with $\Delta_B$ to fit both current data from n-Si/graphene Schottky diodes [48], [49] and values of common-base current gain α from experimental GBTs [14,15]. The same diode current can be fitted with different combinations of the parameters $(h_B, \Delta_B)$, corresponding to different α values for the GBHT: in Fig. 19a the current of almost-ideal diodes from [48] was fitted. The resulting $f_T$ for each α value for a Si GBHT are shown in Fig. 19b. Assuming a common-base current gain of 0.4 (as reported in [15] for GBTs) the resulting cut-off frequency is not higher than 10 GHz. This stresses how critical it is to have a high-quality graphene-semiconductor interface.

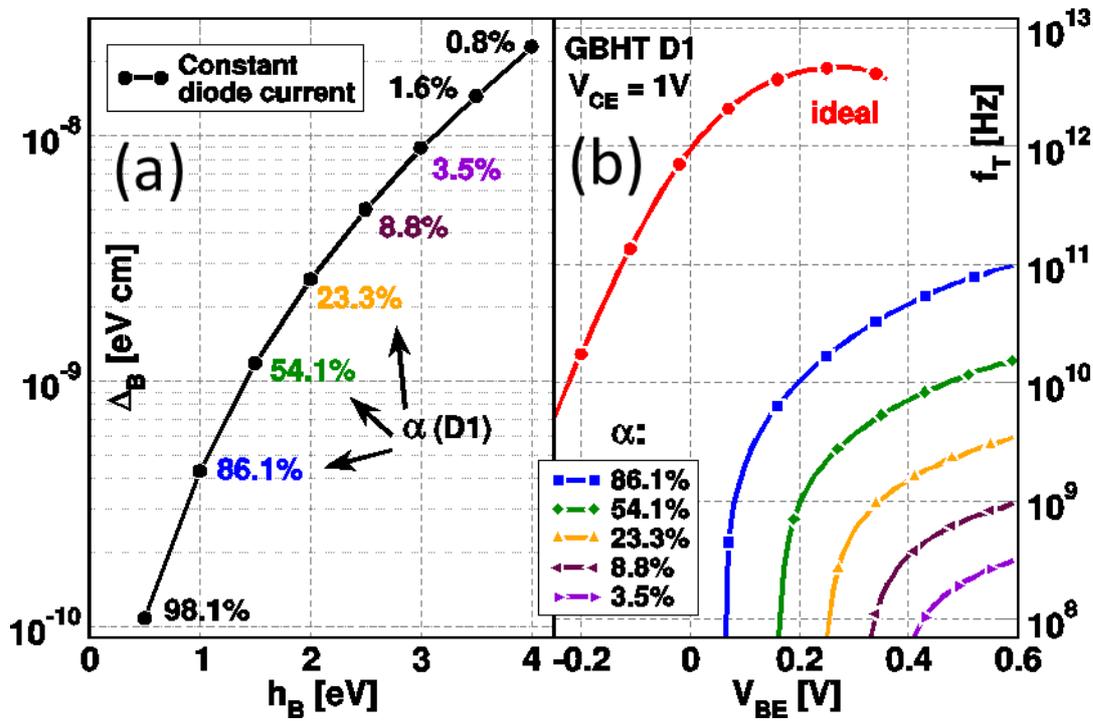

Figure 19 (a) set of the $(h_B, \Delta_B)$ parameter pairs leading to the same simulated I-V curve of the graphene/Si Schottky diode and corresponding to different α values for a typical Si GBHT. (b) cut-off frequency for the GBHT with ideal and non-ideal graphene-Si interface.

Kong et al. reported a simulation study on GBTs based on 2D crystal heterostructures [18]. The authors suggest using 2D crystal materials with appropriate band gaps as the EBI tunneling barrier to exploit their ultimate thickness control and lateral uniformity to prevent current crowding and device to device variability. For instance, hexagonal boron nitride and $MoS_2$ were considered as good candidates as the EBI tunneling barrier. At the collector side, an n-type semiconductor was suggested to form a low Schottky barrier in contact with graphene. This device has been predicted to operate with cutoff frequencies well above 1 THz. However, at this point the technological challenges to realize high performance GBTs or GBHTs based entirely on 2D materials in a manufacturable way by far surpass the likewise considerable challenges of graphene integration into existing silicon technology.

**Conclusions**

This paper reviews the experimental and theoretical state-of-the-art in vertical hot electron transistors with graphene base contacts, GBTs and GBHTs. Simulations predict the performance of these devices surpassing 1 THz in $f_T$ and $f_{max}$. In parallel, early experimental demonstrators show the general feasibility and functionality, without reaching such impressive numbers. Nevertheless, the experimental data has been used to update and improve the models, and even as more and more realistic assumptions are made, the potential for THz performance remains high. Other challenges lie in the integration and manufacturability of these graphene base devices. Among them are the deposition of high quality dielectric films on graphene, low resistance, CMOS compatible electrical contacts, controlled large scale fabrication of graphene and subsequent transfer, low temperature graphene growth directly on desired substrates or contamination issues. These challenges are not unique to GBTs or GBHTs, but must be generally solved by the community, if graphene electronics are to become a reality. In this context, ballistic graphene hot electron transistors remain exciting devices to justify research efforts in this direction.

**Acknowledgements**

The authors wish to dedicate this paper to the memory of Wolfgang Mehr, the inventor of the GBT. The authors thank the IHP cleanroom staff and the technology team for their excellent support and discussions. Support from the European Commission through a European FP7 Project (GRADE, 317839), an ERC Grant (InteGraDe, No. 307311) as well as the German Research Foundation (DFG, LE 2440/1-1) is gratefully acknowledged.